\documentclass[twocolumn,secnumarabic,amssymb,nobibnotes,aps,superscriptaddress,pra]{revtex4-2} 
\usepackage[dvips]{graphicx}
\usepackage{bm}% bold math
\usepackage{latexsym} 
\usepackage{amsmath}
\usepackage{amssymb}
\usepackage{color}

\newcommand{\ui}{{\rm i}}
\newcommand{\veps}{{\varepsilon}}
\newcommand{\eps}{{\epsilon}}

\newcommand{\matq}{{\mathsf{q}}}
\newcommand{\matp}{{\mathsf{p}}}

\newcommand{\bmr}{{\bm r}}
\newcommand{\bmp}{{\bm p}} 
\newcommand{\bmk}{{\bm k}}
\newcommand{\bmK}{{\bm K}}

\newcommand{\bmq}{{\bm q}}

\newcommand{\bmv}{{\bm v}}

\newcommand{\bmE}{{\bm E}}
\newcommand{\bmP}{{\bm P}}

\newcommand{\kB}{k_{\rm B}}

\makeatletter
\renewcommand*{\p@subsection}{}
\renewcommand*{\p@subsubsection}{}
\makeatother

\begin{document}

\title{Ginzburg-Landau action and polarization current in an excitonic insulator model of electronic ferroelectricity}

\author{Hiroto Adachi}
\affiliation{Research Institute for Interdisciplinary Science, Okayama University, Okayama 700-8530, Japan}
%\affiliation{Department of Physics, Okayama University, Okayama 700-8530, Japan}

\author{Naoshi Ikeda} 
\affiliation{Department of Physics, Okayama University, Okayama 700-8530, Japan}

\author{Eiji Saitoh} 
\affiliation{Department of Applied Physics, The University of Tokyo, Hongo, Bunkyo, Tokyo 113-8656, Japan}
\affiliation{Institute of AI and Beyond, The University of Tokyo, Tokyo 113-8656, Japan}
\affiliation{Advanced Institute for Materials Research, Tohoku University, Sendai 980-8577, Japan}
\affiliation{Advanced Science Research Center, Japan Atomic Energy Agency, Tokai 319-1195, Japan}

\date{\today}

\begin{abstract}
  In comparison to transport of spin polarization in ferromagnets, transport of electric polarization in ferroelectrics remains less explored. Taking an excitonic insulator model of electronic ferroelectricity as a prototypical example, we theoretically investigate the low-energy dynamics and transport of electric polarization by microscopically constructing the Ginzburg-Landau action. We show that, because of the scalar nature of the excitonic order parameter, only the longitudinal fluctuations are relevant to the transport of electric polarization. We also formulate the electric polarization diffusion equation, in which the electric-polarization current is defined purely electronically without recourse to the lattice degrees of freedom. 
\end{abstract} 

\pacs{}

\keywords{} 

\maketitle

%%%%%%%%%%%%%%%%%%%%%%%%%%%%%%%%%%%%
\section{Introduction \label{Sec:I}}
%%%%%%%%%%%%%%%%%%%%%%%%%%%%%%%%%%%%

Since the demonstrations of spin transport through an insulating ferromagnet~\cite{Kajiwara10,Uchida10}, much attention has been focused on the spin transport in magnetic insulators~\cite{Wu-Hoffmann13,Chumak15,Brataas20}. Because there are no conduction electrons in magnetic insulators, the spins are not transported by individual excitations of electrons, but instead are carried by collective spin excitations called spin waves or magnons. These collective excitations are, however, not limited to magnetically ordered states. In general, a broken-symmetry state is accompanied by the corresponding collective excitations, and a ferroelectric state is another archetype of such broken-symmetry states~\cite{Anderson-text,Landau-Lifshitz,Chandra-review}. Therefore, it is natural to expect that a novel transport phenomenon can be observed in the ferroelectric state. 

Recently, in a series of publications~\cite{Bauer21,Tang22a,Bauer22,Tang22b} the transport of electric polarization in a displacive ferroelectric material has been discussed extensively. It is then argued that the electric polarization is carried by collective excitations of the ferroelectrically ordered state, termed ``ferrons''. Then, the polarization current, or ``ferron'' current, is defined in terms of the ion displacement field or the phonon operators, because the main focus of those papers is on the ferroelectricity of the displacive type. Conceptually, however, in such a case it is rather hard to distinguish between the ``ferron'' current and the phonon current, since the phonon excitations are simultaneously the electric-polarization excitations in the displacive ferroelectric materials~\cite{Strukov-text}. 

In literature, there is another type of ferroelectricity other than the displacive ferroelectricity, known as the electronic ferroelectricity~\cite{Chandra-review}. A possible realization of the electronic ferroelectricity has extensively been investigated in two-dimensional rare-earth ion oxides $R$ Fe$_2$O$_4$ where $R$ denotes rare-earth-metal elements ($R=$ Lu, Y, Yb, etc.)~\cite{Ikeda05,Nagano07,Fujiwara21}. The electronic ferroelectricity is in general caused by a kind of orbital ordering, and this type of ordering is primarily distinct from the phonon degrees of freedom. Therefore, it is possible in this system to define the ``ferron'' current purely electronically, without relying on the phonon operators.

In this paper we focus on the electronic ferroelectricity, and we investigate the collective excitations and electric-polarization current in the ordered state. For this purpose, we consider an excitonic insulator model of electronic ferroelectricity~\cite{Batyev80}. The excitonic insulator is an electrically insulating material with spontaneously formed bound pairs of electrons and holes, which is otherwise a semimetal or narrow gap semiconductors~\cite{Keldysh65,Jerome67,Halperin68}. When the electron orbital and the hole orbital have opposite parities and when they hybridize, the excitonic insulator possesses a spontaneous polarization, which results in the electronic ferroelectricity~\cite{Batyev80,Portengen96a,Portengen96b}. Specifically, we follow the approach of Ref.~\cite{Portengen96b} and consider the Falicov-Kimball Hamiltonian~\cite{Falicov69,Ramirez70,Batista02} defined on a cubic lattice. Then, utilizing the functional integral methods~\cite{Popov-text}, we microscopically construct the Ginzburg-Landau (GL) action and investigate the resultant collective excitations. Moreover, we discuss how to define the electric-polarization current, or ``ferron'' current, without recourse to the phonon degrees of freedom in the case of electronic ferroelectricity. 

The plan of this paper is as follows. In Sec.~\ref{Sec:II}, we define our model and present the procedure to microscopically construct the GL action. In Sec.~\ref{Sec:III}, we show analytic expression for each term of the GL action. In Sec.~\ref{Sec:IV}, we analyze the GL action in the mean-field approximation as well as in the Gaussian approximation, by using numerical calculations. In Sec.~\ref{Sec:V}, we define the electric polarization in our model, and we investigate the dynamics and transport of electric polarization. Finally, in Sec.~\ref{Sec:VI}, we discuss and summarize our results. We use the units $\hbar=\kB=c=1$ throughout this paper.

%%%%%%%%%%%%%%%%%%%%%%%%%%%%%%%%%%%%%%%%%%%%%%%
\section{Model \label{Sec:II}}  
%%%%%%%%%%%%%%%%%%%%%%%%%%%%%%%%%%%%%%%%%%%%%%%

Following Ref.~\cite{Portengen96b}, we begin with the spinless Falicov-Kimball Hamiltonian: 
%%%
\begin{equation}
  {\cal H} = {\cal H}_0 + {\cal H}_1, \label{eq:FK01}
\end{equation}
%%%
where the first term on the right-hand side, 
%%% 
\begin{equation}
  {\cal H}_0 =
  \sum_\bmp 
    \begin{pmatrix}
      d^\dag_\bmp, & f^\dag_\bmp
    \end{pmatrix}  
  \begin{pmatrix}
    \eps_\bmp & V_\bmp \\
        {V}^*_\bmp & E_0 
      \end{pmatrix}
      \begin{pmatrix}
        d_\bmp \\
        f_\bmp 
      \end{pmatrix}, 
      \label{eq:H0}
\end{equation}
%%%
describes the quadratic part of the Hamiltonian. Here, $d^\dag_\bmp$ creates a $d$-electron with momentum $\bmp$ and energy $\eps_\bmp$, whereas $f^\dag_\bmp$ creates an $f$ electron of momentum $\bmp$ and energy $E_0$. The off-diagonal matrix element $V_\bmp$ is the hybridization between $d$- and $f$-electrons, and $V^*_\bmp$ means the complex conjugate of $V_\bmp$. The chemical potential is set to zero, and all energies are measured from the chemical potential. The second term,
%%%
\begin{equation}
  {\cal H}_1 = \frac{U}{N} \sum_{\bmk} n^{(d)}_{-\bmk} n^{(f)}_{\bmk}, 
  \label{eq:H1}
\end{equation}
%%%
describes the on-site Coulomb repulsion between the $d$- and $f$-electrons, where $n^{(d)}_\bmk = \sum_\bmp d^\dag_\bmp d_{\bmp+ \bmk}$, $n^{(f)}_\bmk = \sum_\bmp f^\dag_\bmp f_{\bmp+ \bmk}$, and $N$ is the number of lattice sites. Note that, for the excitonic insulator phase to simultaneously possess the ferroelectric order, the ferroelectric order parameter must be formed within the same spin component. Therefore, the up and down spin components are completely decoupled in our approach, and we can safely use the spinless model by absorbing the effect of two spin components into the spin degeneracy factor. When there exists a finite spin-orbit interaction, although spin is no longer a good quantum number, the spin index is replaced by the ``pseudospin'' index which labels the Kramers doublet (pseudospin up and down) because of the time-reversal symmetry~\cite{Anderson84}. Therefore, the present approach remains unchanged qualitatively even in the presence of a sizable spin-orbit interaction. 

The Falicov-Kimball model [Eq.~(\ref{eq:FK01})] is known to possess the $d$-$f$ excitonic insulator phase as a mean-field solution~\cite{Khomskii76,Kanda76,Leder78a}. As shown below, in the presence of the inter-orbital hybridization, the excitonic insulator phase simultaneously possesses the electronic polarization. Applicability of the electronic ferroelectricity model to mixed-valence compounds has been discussed~\cite{Portengen96a,Portengen96b}, ranging from  SmB$_6$~\cite{Wachter85} and TmSe$_{0.45}$Te$_{0.55}$~\cite{Bucher91} to Sm$_3$Se$_4$~\cite{Goto93}. Although the existence of the excitonic insulator phase in this model has been challenged by several authors~\cite{Czycholl99,Farkasovsky99,Zlatic01,Batista02}, we respect the fact in this work that this is a minimal model system that exhibits the electronic ferroelectricity~\cite{Batyev80,Portengen96a,Portengen96b}. We therefore start from this model, and pursue the consequence of the emergence of electronic ferroelectricity. Following this philosophy, we first change the fermion ordering in ${\cal H}_1$ and rewrite the Hamiltonian in the following way, 
%%%
\begin{equation}
  %  {\cal H}_1
  {\cal H}
  = 
  {\cal H}_0 -\frac{U}{N} \sum_\bmq B^\dag_\bmq B_\bmq,
  \label{eq:H_exciton01}
\end{equation}
%%%
where we have defined 
%%%
\begin{equation}
  B_\bmq = \sum_\bmp f^\dag_\bmp d_{\bmp+ \bmq}, 
\end{equation}
%%%
and we have disregarded terms which can be absorbed in the chemical potential shift. Note that we ignore phonons here. Howewer, in the candidate excitonic insulator Ta$_2$NiSe$_5$ the active role of phonons has been pointed out~\cite{Chen-arXiv}, and ignoring influences of phonons on the exitonic insulator phase in a real material could be an oversimplification. Note also that, as we will see below, the key ingredients in our model Halmiltonian~(\ref{eq:FK01}) are as follows: (i) two orbitals ($d$ and $f$) with opposite parities, and (ii) a nonzero hybridization $V_\bmp$ between the two orbitals, both of which are important to stabilize the electronic ferroelectricity phase. Concerning the latter point, if we take SmB$_6$ as an example, the magnitude of $V_\bmp$ is estimated to be of the order of ten to hundreds of meV~\cite{Frantzeskakis13}.

In the following, we use the functional integral representation of the problem~\cite{Popov-text} and construct the GL action. We first define the two-component field $\psi$ by 
%%%
\begin{equation}
  \psi =  \begin{pmatrix}
        d_\bmp \\
        f_\bmp 
      \end{pmatrix}, 
\end{equation}
%%%
and we represent ${\cal H}_0$ as 
%%% 
\begin{eqnarray}
  {\cal H}_0 &=&
  \sum_\bmp \psi^\dag_\bmp \widehat{h}_0(\bmp) \psi_\bmp , 
  \label{eq:H0b}
\end{eqnarray}
%%%
where we introduced the $2 \times 2$ matrix 
%%%
\begin{equation}
  \widehat{h}_0(\bmp)
  = 
  \begin{pmatrix}
    \eps_\bmp & V_\bmp \\
    {V}^*_\bmp & E_0 
  \end{pmatrix}. 
\end{equation}
%%%  \label{eq:H_exciton01}
Then, the partition function for the present model [Eq.~(\ref{eq:H_exciton01})] is represented as
%%%
\begin{equation}
  {\cal Z} = \int {\cal D}[\psi, \psi^\dag] e^{-{\cal S}}
  \label{eq:Z01}
\end{equation}
%%%
where the action ${\cal S}$ is defined by 
%%%
\begin{equation}
  {\cal S} =
  \int_0^{1/T} d \tau \Big\{ \sum_\bmp \psi^\dag_\bmp \big( \partial_\tau + \widehat{h}_0(\bmp) \big) \psi_\bmp
  - \frac{U}{N} \sum_\bmq B^\dag_\bmq B_\bmq \Big\} , 
  \label{eq:S01}
\end{equation}
%%%
and ${\cal D}[\psi,\psi^\dag]$ in Eq.~(\ref{eq:Z01}) denotes the functional integral over $\psi$ and $\psi^\dag$, with an understanding that $\psi$ and $\psi^\dag$ represent the Grassman variables when used within the functional integral~\cite{Coleman-text}. 

Our next step is to use the Hubbard-Stratonovich transformation to obtain the effective action. To do so, we insert the identity 
%%%
\begin{equation}
  1 = \int {\cal D}[\zeta, \zeta^*] 
  \exp \left(- \frac{1}{U} \sum_\bmq \int_0^{1/T} d \tau \zeta^*_\bmq \zeta_\bmq \right) 
\end{equation}
%%%
into Eq.~(\ref{eq:Z01}), where the normalization factor is understood to have been absorbed into the integration measure over the auxiliary field $\zeta$. Then, after a shift transformation, 
%%%
\begin{equation}
  \zeta_\bmq = \Delta_\bmq+  \frac{U}{\sqrt{N}} B_\bmq,
  \label{eq:Stratnovich01}
\end{equation}
%%% 
the partition function can be written as
%%%
\begin{equation}
  {\cal Z} = \int {\cal D}[\Delta, \Delta^*]
  e^{-\frac{1}{U} \int_0^{1/T} d \tau \sum_\bmq \Delta^*_\bmq \Delta_\bmq }
  \int {\cal D}[\psi, \psi^\dag] e^{-{\cal S'}},
  \label{eq:Z02}
\end{equation}
%%%
where ${\cal S'}$ is given by
%%%
\begin{equation}
  {\cal S'} = \int_0^{1/T} d \tau \sum_{\bmp,\bmp'}
  \psi^\dag_\bmp \Big( [\partial_\tau + \widehat{h}_0(\bmp)] \delta_{\bmp,\bmp'}
  + \widehat{h}_1(\bmp,\bmp') \Big) \psi_{\bmp'},
  \label{eq:S02}
\end{equation}
%%%
and the $2 \times 2$ matrix $\widehat{h}_1$ is defined by
%%%
\begin{equation}
  \widehat{h}_1 (\bmp,\bmp')
  =
  \frac{1}{\sqrt{N}}\begin{pmatrix}
    0 & \Delta_{\bmp-\bmp'} \\
    \Delta^*_{\bmp'-\bmp} & 0 
  \end{pmatrix}. 
\end{equation}
%%%
The action in Eq.~(\ref{eq:S02}) is quadratic with respect to the fermion field $\psi$, and hence the Gaussian integral over $\psi$ can be performed. This procedure yields the effective action represented solely by the order-parameter field $\Delta$, 
%%%
\begin{equation}
  {\cal Z} = {\cal Z}_0 \times \int {\cal D}[\Delta, \Delta^*]
  \exp \{-{\cal S}_{\rm eff} \},
  \label{eq:S_eff01}
\end{equation}
%%%
where ${\cal Z}_0$ is the partition function in the absence of $\Delta$. The effective action ${\cal S}_{\rm eff}$ is then given by 
%%%
\begin{equation}
  {\cal S}_{\rm eff} = \int_0^{1/T} d \tau \sum_\bmq \frac{1}{U} \Delta^*_\bmq \Delta_\bmq
  + {\rm Tr} \sum_{l=1}^\infty \frac{1}{l} \left( \widehat{G} \widehat{h}_1 \right)^l
  \label{eq:S03}
\end{equation}
%%%
where $\widehat{G}= -\left( \partial_\tau + \widehat{h}_0 \right)^{-1}$ is the Green's function. The effective action ${\cal S}_{\rm eff}$ can be expanded in powers of $\Delta$, 
%%%
\begin{equation}
  {\cal S}_{\rm eff} =   \sum_{l=1}^\infty {\cal S}_{l}, 
  \label{eq:S04}
\end{equation}
%%%
where the index $l$ in ${\cal S}_l$ represents the power with respect to $\Delta$.

%%%%%%%%%%%%%%%%%%%%%%%%%%%%%%%%%%%%
\section{Ginzburg-Landau action \label{Sec:III}} 
%%%%%%%%%%%%%%%%%%%%%%%%%%%%%%%%%%%%
In this section, focusing on a region near the phase transition, we expand the effective action [Eq.~(\ref{eq:S04})] up to the quartic term with respect to $\Delta$. In the absence of an external electric field, the resultant GL action has terms of the second and fourth orders as follows: 
%%%
\begin{equation}
  {\cal S}_{\rm GL} = {\cal S}_2 + {\cal S}_4, 
  \label{eq:S_GL01}
\end{equation}
%%%
where higher-order terms are disregarded. Note that, because of the two-band nature of the system under discussion, this GL action has more terms than a single-band case~\cite{Popov-text}. 

In the following calculations, it is convenient to work with the Matsubara frequency representation, such that we introduce the expansion 
%%%
\begin{equation}
  \Delta_{\bmq}(\tau) = \sqrt{T} \sum_{\ui \omega_m} \Delta_{\bmq,\ui \omega_m} e^{-\ui \omega_m \tau}, 
\end{equation}
%%%
where $\ui= \sqrt{-1}$, and $\omega_m = 2 \pi T m$ is a bosonic Matsubara frequency with integer $m$. Besides, the Green's function $\widehat{G}$ has the following representation: 
%%%
\begin{eqnarray}
  \widehat{G}  (\bmp, \ui \veps_n) &=&
  \begin{pmatrix} 
    G_1(\bmp, \ui \veps_n)   & G_2 (\bmp, \ui \veps_n)  \\
    G_3 (\bmp, \ui \veps_n)  & G_4 (\bmp, \ui \veps_n) 
  \end{pmatrix}  \nonumber \\
  &=&
  \frac{1}{ D(\bmp, \ui \veps_n)}
  \begin{pmatrix}
    \ui \veps_n - E_0 & V_\bmp \\
    V^*_\bmp  &  \ui \veps_n - \eps_\bmp 
  \end{pmatrix},
  \label{eq:Gfunc01}
\end{eqnarray}
%%%
where $ D(\bmp, \ui \veps_n)= (\ui \veps_n- \lambda_\bmp^+)(\ui \veps_n- \lambda_\bmp^-)$, $\veps_n= 2 \pi T (n+1/2)$ is a fermionic Matsubara frequency with integer $n$, and $\lambda_\bmp^\pm$ is the dispersion determined by ${\rm det}[\lambda_\bmp^\pm- \widehat{h}_0(\bmp)]=0$. More precisely, $\lambda_\bmp^\pm$ have expressions 
%%% 
\begin{eqnarray}
  \lambda_\bmp^\pm &=& Q_\bmp \pm R_\bmp, 
\end{eqnarray}
%%%
where $Q_\bmp$ and $R_\bmp$ are given by 
%%%
\begin{eqnarray}
  Q_\bmp &=& \frac{\veps_\bmp+ E_0}{2}, \\
  R_\bmp &=& \sqrt{\left(\frac{\veps_\bmp- E_0}{2}\right)^2+ |V_\bmp|^2}. 
\end{eqnarray}
%%%

%%%%%%%%%%%%%%%%%%%%%%%%%%%%%%%%
\subsection{Quadratic term}
%%%%%%%%%%%%%%%%%%%%%%%%%%%%%%%%
We first calculate the quadratic term of the action, which has the following representation: 
%%%
\begin{equation}
  {\cal S}_2 = \frac{1}{2} \sum_{\mathsf{q}} \Big( a_1(\mathsf{q}) \Delta^*_{\mathsf{q}} \Delta_{\mathsf{q}} + a_2(\mathsf{q}) \Delta^*_{\mathsf{q}} \Delta^*_{-\mathsf{q}} + a_3(\mathsf{q}) \Delta_{\mathsf{q}} \Delta_{-\mathsf{q}} \Big), 
\end{equation}
%%%
where we have introduced a shorthand notation $\mathsf{q}= (\bmq, \ui \omega_m)$. Here, the coefficient $a_i(\matq)$ ($i=1$, $2$, and $3$) is given by 
%%%
\begin{eqnarray}
  a_1(\matq) &=&
  2 \left( \frac{1}{U}+ \frac{T}{N} \sum_\matp G_1 (\matp+ \matq) G_4 (\matp) \right),
  \label{eq:a1}\\
  a_2(\matq) &=&
  \frac{T}{N}  \sum_\matp G_2 (\matp+ \matq) G_2 (\matp),   \label{eq:a2} \\
  a_3(\matq) &=&
  \frac{T}{N}  \sum_\matp G_3 (\matp+ \matq) G_3 (\matp),   \label{eq:a3}
\end{eqnarray}
%%%
where we have defined $\mathsf{p}= (\bmp, \ui \veps_n)$.

Now, following the philosophy of the GL expansion, we focus on small spatial variations as well as slow temporal variations. In the present formulation, this corresponds to an expansion for small $\bmq$ and $\omega$, where $\omega$ is a real frequency after the analytic continuation $\ui \omega_m \to \omega + \ui \delta$ ($\delta= 0^+$). Under the static condition $\omega=0$, expansion with respect to $\bmq$ gives the bare correlation length, 
%%%
\begin{eqnarray}
  a_1(\bmq,0)- a_1({\bm 0},0) 
  &=&
  \xi_1^2 q^2, \\
  a_2(\bmq,0)- a_2({\bm 0},0)
  &=&
  \xi_2^2 q^2, \\
   a_3(\bmq,0)- a_3({\bm 0},0)
  &=&
   \xi_3^2 q^2, 
\end{eqnarray}
%%%
where $\xi_i$ ($i=1$, $2$, and $3$) is given by 
%%%
\begin{eqnarray}
  \xi_1^2 
  &=&
  \frac{2T}{N}  \sum_{\mathsf{p}}
  \frac{ \big[ G_1(\mathsf{p}) \big]^2 G_4(\mathsf{p}) }{3} 
  \Big\{ G_1(\mathsf{p}) \bmv_\bmp^2+ \frac{M_\bmp}{2} \Big\} , \label{eq:xixi01}    \\
  \xi_2^2 
  &=&
  \frac{T}{N}  \sum_{\mathsf{p}}
  \frac{ G_1(\mathsf{p}) \big[ G_2(\mathsf{p}) \big]^2 }{3}
  \Big\{ G_1(\mathsf{p}) \bmv_\bmp^2+ \frac{M_\bmp}{2} \Big\} , \label{eq:xixi02} \\
  \xi_3^2 
  &=&
  \frac{T}{N}  \sum_{\mathsf{p}}
  \frac{ G_1(\mathsf{p}) \big[ G_3(\mathsf{p}) \big]^2 }{3}
  \Big\{ G_1(\mathsf{p}) \bmv_\bmp^2+ \frac{M_\bmp}{2} \Big\} \label{eq:xixi03} , 
\end{eqnarray}
%%%
where $\bmv_\bmp= {\bm \nabla}_\bmp \eps_\bmp$, and $M_\bmp= \sum_{j=x,y,z} \partial^2 \eps_\bmp /\partial p_j^2$. 

Calculation of the dynamical term ($\omega \neq 0$) requires a little more care than the static term. In this case, it is convenient to use the knowledge of the contour integral. Using the correspondence $|\omega_m| \leftrightarrow -\ui \omega$ under the analytic continuation $\ui \omega_m \to \omega + \ui \delta$ ($\delta= 0^+$), we obtain~\cite{AGD} 
%%%
\begin{eqnarray}
  a_1(\bmq,\ui \omega_m)- a_1({\bm 0},0) 
  &=&
  \frac{|\omega_m| }{\Gamma_1(\bmq)}, \label{eq:Im_a1} \\
  a_2(\bmq,\ui \omega_m)- a_2({\bm 0},0) 
  &=&
  \frac{|\omega_m| }{\Gamma_2(\bmq)}, \label{eq:Im_a2} \\  
  a_3(\bmq,\ui \omega_m)- a_3({\bm 0},0)
  &=&
  \frac{|\omega_m| }{\Gamma_3(\bmq)}, \label{eq:Im_a3}
\end{eqnarray}
%%%
where $\Gamma_i(q)$ ($i=1$, $2$, and $3$) is given by
%%%
\begin{eqnarray}
  \frac{1}{\Gamma_1(\bmq)} 
  &=&
  \frac{1}{2\pi T} \int_{-\infty}^\infty   d \veps \; 
  \frac{1} {\cosh^2 \left( \frac{\veps}{2T}\right) } 
  \nonumber \\
  && \times 
  \frac{1}{N}\sum_\bmp  {\rm Im} G^R_1(\bmp+ \bmq, \veps) 
      {\rm Im} G^R_4 (\bmp, \veps),    \label{eq:Gamma01} \\
      \frac{1}{\Gamma_2(\bmq)} 
  &=&
      \frac{1}{4 \pi T} \int_{-\infty}^\infty   {d \veps} \; 
      \frac{1} {\cosh^2 \left( \frac{\veps}{2T}\right) }
        \nonumber \\
  & &\times  
      \frac{1}{N}\sum_\bmp  {\rm Im} G^R_2(\bmp+ \bmq, \veps) 
       {\rm Im} G^R_2 (\bmp, \veps),  \label{eq:Gamma02} \\
       \frac{1}{\Gamma_3(\bmq)} 
  &=&
  \frac{1}{4 \pi T} \int_{-\infty}^\infty d \veps \; 
  \frac{1} {\cosh^2 \left( \frac{\veps}{2T}\right) }
  \nonumber \\
  && \times  
  \frac{1}{N}\sum_\bmp  {\rm Im} G^R_3(\bmp+ \bmq, \veps) 
       {\rm Im} G^R_3 (\bmp, \veps).  \label{eq:Gamma03}  
\end{eqnarray}
%%%
In these equations, $\widehat{G}^R(\bmp,\veps)= \widehat{G}(\bmp,\ui \veps_n)|_{\ui \veps_n \to \veps+ \ui \delta}$ is the retarded component of the Green's function.

%%%%%%%%%%%%%%%%%%%%%%%%%%%%%%%%
\subsection{Quartic term} 
%%%%%%%%%%%%%%%%%%%%%%%%%%%%%%%%
Next, we calculate the quartic term. As mentioned in the beginning of this section, due to the two-band nature of the problem, we find that this term involves several contributions as follows: 
%%%
\begin{eqnarray}
  {\cal S}_4 &=& \frac{T}{4 N}\sum_{\matq_1,\matq_2,\matq_3,\matq_4}
  \delta_{\matq_1+ \matq_2+ \matq_3+ \matq_4,0} \nonumber \\
  &\times & \Big\{ \big( b_1 + b_2 \big) 
  \Delta^*_{-\matq_1} \Delta^*_{-\matq_2} \Delta_{\matq_3} \Delta_{\matq_4} \nonumber \\
  &+& b_3   \Delta^*_{-\matq_1} \Delta_{\matq_2} \Delta_{\matq_3} \Delta_{\matq_4} 
  + b_4   \Delta^*_{-\matq_1} \Delta^*_{-\matq_2} \Delta^*_{-\matq_3} \Delta_{\matq_4} \nonumber \\
  &+&  b_5  \Delta^*_{-\matq_1} \Delta^*_{-\matq_2} \Delta^*_{-\matq_3} \Delta^*_{-\matq_4} 
  + b_6 \Delta_{\matq_1} \Delta_{\matq_2} \Delta_{\matq_3} \Delta_{\matq_4} \Big\}, \nonumber \\
\end{eqnarray}
%%%
where each coefficient $b_i$ ($i=1,2, \cdots, 6$) is given by 
%%%
\begin{eqnarray}
  b_1 &=&
  2 \frac{T}{N}  \sum_\matp \big( G_1 (\matp) \big)^2 \big( G_4 (\matp) \big)^2, \\
  b_2 &=&
  4 \frac{T}{N}  \sum_\matp G_1 (\matp) G_2 (\matp) G_3 (\matp) G_4 (\matp), \\
  b_3 &=&
  4 \frac{T}{N}  \sum_\matp G_1 (\matp) \big( G_3 (\matp) \big)^2 G_4 (\matp), \\
  b_4 &=&
  4 \frac{T}{N}  \sum_\matp G_1 (\matp) \big( G_2 (\matp) \big)^2 G_4 (\matp), \\
  b_5 &=&
  \frac{T}{N}  \sum_\matp \big( G_2 (\matp) \big)^4, \\
  b_6 &=&
  \frac{T}{N}  \sum_\matp \big( G_3 (\matp) \big)^4. 
\end{eqnarray}
%%%

In the next section, the GL action obtained above is analyzed using the mean-field approximation as well as the Gaussian approximation.

%%%%%%%%%%%%%%%%%%%%%%%%%%%%%%%%%%%%%%%%%%%%%
\section{Mean-field solution and Gaussian fluctuations \label{Sec:IV}} 
%%%%%%%%%%%%%%%%%%%%%%%%%%%%%%%%%%%%%%%%%%%%%

In this section, we first obtain the mean-field solution for the GL action [Eq.~(\ref{eq:S_GL01})] and then investigate the Gaussian fluctuations around the mean-field solution. Next, we perform a model calculation to evaluate GL coefficients. The approach used in this work, i.e., the mean-field approximation and the Gaussian fluctuations above the mean-field solution, is a standard technique to investigate the collective excitations in the broken-symmetry state (see Chap.~6 of Ref.~\cite{Altland-Simons}). Note that we are not dealing with the critical phenomena. Although within the Ginzburg temperature region the non-Gaussian fluctuations neglected in our approach become important to the critical phenomena, the present approach provides a consistent description of the phase transition outside the Ginzburg temperature region~\cite{Chaikin-Lubensky}. Needless to say, in low-dimensional systems with extraordinary strong fluctuations, the validity of the mean-field approximation could break down.

%%%%%%%%%%%%%%%%%%%%%%%%%%%%%%%%%%%%%%%%%%%%%%%%
\subsection{Construction of the Gaussian action}
%%%%%%%%%%%%%%%%%%%%%%%%%%%%%%%%%%%%%%%%%%%%%%%%
The ground state of the GL action [Eq.~(\ref{eq:S_GL01})] is spatially uniform because the gradient energy is positive in our model calculation as is shown in the next subsection. Under this condition, we substitute into Eq.~(\ref{eq:S_GL01}) the following decomposition~\cite{Popov-text} 
%%%
\begin{equation}
  \Delta_{\mathsf q} = \sqrt{N/T} \delta_{{\mathsf q},0} \Delta_0 + \delta \Delta_{\mathsf q},
  \label{eq:Gapdecomp01}
\end{equation}
%%% 
where $\Delta_0$ is the mean-field solution determined by the static saddle-point condition, whereas $\delta \Delta_{\mathsf q}$ are the fluctuations. Note that when we consider a nearest-neighbor hybridization between two orbitals with opposite parities, i.e., $d$- and $f$-electrons, $V_\bmp$ is purely imaginary and odd in $\bmp$~\cite{Sandu05}, from which it follows that $\Delta_0$ becomes purely real~\cite{Portengen96b}. In accordance with Eq.~(\ref{eq:Gapdecomp01}), we decompose the GL action up to the second order with respect to the fluctuations, 
%%%
\begin{equation}
  {\cal S}_{\rm GL} = {\cal S}_0 + {\cal S}_{\rm GAUSS}, 
  \label{eq:S_GL02}
\end{equation}
%%%
where ${\cal S}_0$ is the zeroth order in $\delta \Delta_{\mathsf q}$, whereas ${\cal S}_{\rm GAUSS}$ is the second order. 

The first term on the right-hand side of Eq.~(\ref{eq:S_GL02}) disregards the fluctuations of $\Delta$ and leads to the free energy in the mean-field approximation, 
%%%
\begin{equation} 
  {\cal S}_0 = \frac{F_{\rm MFA}}{T}, 
  \label{eq:S0a}
\end{equation}
%%%
where the mean-field free energy takes the form 
%%%
\begin{equation}
  F_{\rm MFA} = N \left( \frac{a_0}{2} \Delta_0^2 + \frac{b_0}{4} \Delta_0^4 \right), 
\end{equation}
%%%
where
%%%
\begin{eqnarray}
  a_0 &=& a_1({\mathsf 0})+a_2({\mathsf 0})+a_3({\mathsf 0}), \label{eq:a0}\\
  b_0 &=& \sum_{i=1}^6 b_i \label{eq:b0}.
\end{eqnarray}
%%%
The phase transition into the ferroelectric phase is signified by the condition $a_0 \leq 0$. Here the amplitude of the condensate is determined by the static saddle-point condition of $F_{\rm MFA}$ in Eq.~(\ref{eq:S0a}), namely, 
%%%
\begin{equation}
  \Delta_0^2 = \frac{|a_0|}{b_0}.
  \label{eq:Delta0}
\end{equation}
%%%

The second term on the right-hand side of Eq.~(\ref{eq:S_GL02}) is the Gaussian action for the electronic ferroelectricity phase. Note that the term linear in $\delta \Delta_{\mathsf{q}}$ vanishes because of the saddle-point condition [Eq.~(\ref{eq:Delta0})]. The Gaussian action has the form, 
%%%
\begin{equation}
  {\cal S}_{\rm GAUSS} =
  \sum_{\mathsf{q}} 
  \begin{pmatrix}
    \delta \Delta^*_{\mathsf{q}}, & \delta \Delta_{-\mathsf{q}}
  \end{pmatrix} 
  \begin{pmatrix}
    A(\mathsf{q}) & B(\mathsf{q}) \\
    B^*(\mathsf{q}) & A(\mathsf{q})
  \end{pmatrix}
  \begin{pmatrix}
    \delta \Delta_{\mathsf{q}} \\
    \delta \Delta^*_{-\mathsf{q}} 
  \end{pmatrix}
\end{equation}
%%%
where each matrix element is given by 
%%%
\begin{eqnarray}
  A(\mathsf{q}) &=& \frac{a_1(\mathsf{q})}{4}+ \frac{4b_1+ 4b_2+ 3b_3+ 3b_4}{8} \Delta_0^2 , \\
  B(\mathsf{q}) &=& \frac{a_2(\mathsf{q})}{2}+ \frac{b_1+ b_2+ 3b_4+ 6b_5}{4} \Delta_0^2 , 
\end{eqnarray}
%%%
and we use a property $a_i(-\mathsf{q})= a_i(\mathsf{q})$ for $i= 1$, $2$, and $3$. The Gaussian action can be diagonalized by introducing the new fields $\alpha_{\mathsf{q}}$ and $\beta_{\mathsf{q}}$: 
%%%
\begin{equation}
  \begin{pmatrix}
    \delta \Delta_\mathsf{q} \\
    \delta \Delta^*_{-\mathsf{q}}
  \end{pmatrix}
  =
  \frac{1}{\sqrt{2}}
  \begin{pmatrix}      
    1 & 1 \\
    -1 & 1 
  \end{pmatrix}
  \begin{pmatrix}
    \alpha_\mathsf{q} \\
    \beta_{\mathsf{q}}
  \end{pmatrix}, 
  \label{eq:unitaryT01}
\end{equation}
%%%
which transforms the Gaussian action into 
%%%
\begin{equation}
  {\cal S}_{\rm GAUSS} =
  \sum_{\mathsf{q}} \Big( \Omega_- (\mathsf{q}) \alpha^*_{\mathsf{q}} \alpha_{\mathsf{q}}
  + \Omega_+ (\mathsf{q}) \beta^*_{\mathsf{q}} \beta_{\mathsf{q}}\Big),
  \label{eq:S_GL03}
\end{equation}
%%%
where
%%%
\begin{eqnarray}
  \Omega_- (\mathsf{q}) &=& A(\mathsf{q})- B(\mathsf{q}), \\
  \Omega_+ (\mathsf{q}) &=& A(\mathsf{q})+ B(\mathsf{q}).
\end{eqnarray}
%%%

If we invert Eq.~(\ref{eq:unitaryT01}), we obtain  
%%%
\begin{equation}
  \begin{pmatrix}
    \alpha_{\mathsf{q}} \\
    \beta_{\mathsf{q}}
  \end{pmatrix}
  =
  \frac{1}{\sqrt{2}}
  \begin{pmatrix}      
    \delta \Delta_{\mathsf{q}}- \delta \Delta^*_{-\mathsf{q}} \\
    \delta \Delta_{\mathsf{q}}+ \delta \Delta^*_{-\mathsf{q}} 
  \end{pmatrix}, 
  \label{eq:unitaryT02}
\end{equation}
%%%
from which we see that $\alpha_{\mathsf{q}}$ corresponds to the phase mode of the complex order parameter fluctuations, whereas $\beta_{\mathsf{q}}$ corresponds to the amplitude mode~\cite{Nagaosa-text}. In accordance with the small $\omega_m$ expansion in Eqs.~(\ref{eq:Im_a1})-(\ref{eq:Im_a3}), the eigen value $\Omega_\pm(\mathsf{q})$ can be decomposed as follows: 
%%%
\begin{equation}
  \Omega_\pm (\bmq, \ui \omega_m)= \Lambda_\pm (\bmq)+ \frac{|\omega_m|}{\Gamma_\pm (\bmq)},
  \label{eq:Omega01}
\end{equation}
%%%
where $\Lambda_\pm (\bmq)$ is defined by 
%%%%
\begin{equation}
  \Lambda_\pm (\bmq) = \Omega_\pm (\bmq, \ui \omega_m = 0).
  \label{eq:Lambda01}
\end{equation}
%%%

%%%%%%%%%%%%%%%%%%%%%%%%%%%%%%%%%%%%%%%%%%%%%%%%%
\subsection{Numerical calculation}
%%%%%%%%%%%%%%%%%%%%%%%%%%%%%%%%%%%%%%%%%%%%%%%%%

%%%%%%%%%%%%%%%%%%%%%%%%%%%%%%%%%%%%% 
\begin{figure}[t] 
  \begin{center}
    \includegraphics[width=8cm]{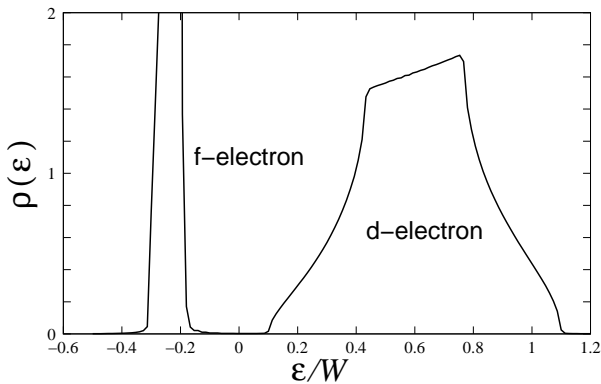}
  \end{center}
  \caption{An example of the density of states considered in this work as a function of energy in the non-interacting case ($U=0$) for $\eps_{\rm min}/W= 0.1$, $V/W= 0.1$, and $E_0/W= -0.2$, where $W$ is the $d$-electron bandwidth}.
  \label{fig:DOS01}
\end{figure}
%%%%%%%%%%%%%%%%%%%%%%%%%%%%%%%%%%%%

In this subsection, we show results of our numerical evaluation for the GL action. For simplicity we assume a three-dimensional cubic lattice, and consider the nearest-neighbor contributions to the hopping and the hybridization. Then, for the single-particle Hamiltonian ${\cal H}_0$ [Eq.~(\ref{eq:H0})], we use 
%%%
\begin{equation}
  \eps_\bmp = \eps_{\rm min}+ \left( \frac{W}{6} \right) \left[ 3- \sum_{\nu= x,y,z} \cos(p_\nu a) \right] 
\end{equation}
%%%
and
%%%
\begin{equation}
  V_\bmp = \ui V \sum_{\nu= x,y,z} \sin(p_\nu a), 
\end{equation}
where $a$ is the lattice spacing, and as mentioned before the hybridization $V_\bmp$ is purely imaginary and odd in $\bmp$~\cite{Portengen96b,Sandu05}. In our numerical calculation, we use a cubic mesh of $50 \times 50 \times 50$ in the Brillouin zone (except for Fig.~\ref{fig:DOS01} which requires a $500 \times 500 \times 500$ mesh for a fine spectral resolution), and we consider $2 N_{\rm max}+1$ points in the Matsubara sum, where we take $N_{\rm max} = E_{\rm max}/T$ with $E_{\rm max}= 10W$. Note that, in our numerical calculation, length is measured in unit of the lattice spacing $a$, and all energies are measured in units of the $d$-electron bandwidth $W$.

Figure~\ref{fig:DOS01} shows the density of states~\cite{Leder78b} 
%%%
\begin{equation}
  \rho(\veps)=
  %  - \left. \frac{1}{\pi N} {\rm Im} \sum_\bmp \sum_{j=1,4} G_j(\bmp, \ui \veps_n )
  - \left. \frac{1}{\pi N} {\rm Im} \sum_\bmp {\rm Tr} \; \widehat{G}(\bmp, \ui \veps_n )
  \right|_{\ui \veps_n \to \veps+ \ui \eta}, 
\end{equation}
%%%
calculated for a system without the Coulomb interaction ${\cal H}_1$ (i.e., $U=0$), where $\eps_{\rm min}/W= 0.1$, $V/W= 0.1$, $E_0/W= -0.2$, and $\eta/W= 0.001$ are used. As stated before, all energies are measured from the chemical potential. The upper band at higher energies with a wider bandwidth $W$ is mainly formed by $d$-electrons. By contrast, the lower band centered around $\veps= E_0$ is mainly formed by $f$-electros, which is slightly broadened because of the finite hybridization $V_\bmp$. Throughout this work, we assume that the system possesses a gap of the order of $0.1$, as shown in Fig.~\ref{fig:DOS01}. Note that in the spinless two-orbital Falicov-Kimball model under consideration, one lattice site can host two electrons in $d$- and $f$-orbitals. In this work, we consider a half filled case as seen in Fig.~\ref{fig:DOS01}. 

%%%%%%%%%%%%%%%%%%%%%%%%%%%%%%%%%%%%% 
\begin{figure}[t] 
  \begin{center}
    \includegraphics[width=8cm]{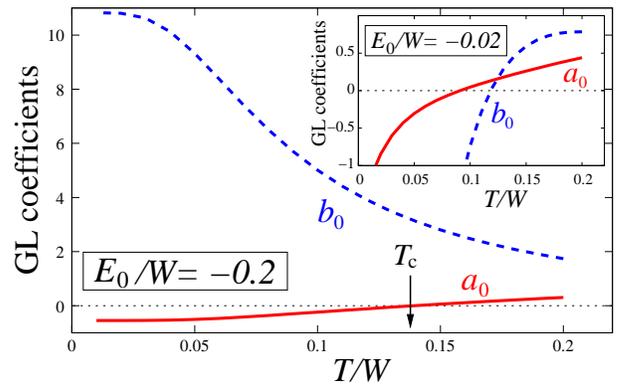}
  \end{center}
  \caption{GL coefficients $a_0$ (solid red line) and $b_0$ (dashed blue line) for $\eps_{\rm min}/W= 0.1$, $V/W= 0.1$, $U/W=1.0$, and $E_0/W= -0.2$. The inset shows the result for $E_0/W= -0.02$ while keeping other parameters the same as those in the main panel.}
  \label{fig:GLab01}
\end{figure}
%%%%%%%%%%%%%%%%%%%%%%%%%%%%%%%%%%%%

In Fig.~\ref{fig:GLab01}, we show the temperature dependence of the GL coefficients $a_0$ [Eq.~(\ref{eq:a0})] and $b_0$ [Eq.~(\ref{eq:b0})]. For parameters $\eps_{\rm min}/W=0.1$, $V/W=0.1$, $U/W=1.0$, and $E_0/W= -0.2$, the quadratic coefficient $a_0$ changes sign from positive to negative upon lowering the temperature, signifying a phase transition at $T_{\rm c}/W= 0.14$. At the same time, the quartic coefficient $b_0$ is positive in this temperature region, confirming the second-order phase transition at $T_{\rm c}$. However, as shown in the inset of Fig.~\ref{fig:GLab01}, when the $f$-level energy is changed from $E_0/W=-0.2$ to $E_0/W= -0.02$ while keeping other parameters unchanged, the quartic coefficient $b_0$ becomes negative upon cooling before the quadratic coefficient $a_0$ becomes negative, meaning that the transition is of the first order.

%%%%%%%%%%%%%%%%%%%%%%%%%%%%%%%%%%%%% 
\begin{figure}[t] 
  \begin{center}
    \includegraphics[width=7.5cm]{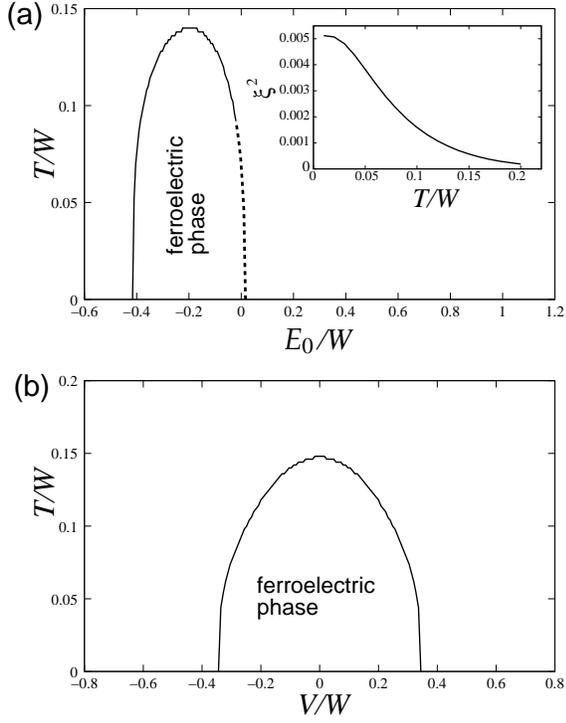}
  \end{center}
  \caption{(a) $T$-$E_0$ phase diagram in the mean-field approximation for $\eps_{\rm min}/W= 0.1$, $V/W= 0.1$, and $U/W=1.0$. The solid line represents the second-order transition line, whereas the dashed line denotes the first-order transition line. Inset: Temperature dependence of $\xi^2$ [Eq.~(\ref{eq:xixi})] for $\eps_{\rm min}/W= 0.1$, $V/W= 0.1$, $E_0/W= -0.2$, and $U/W=1.0$. (b) $T$-$V$ phase diagram in the mean-field approximation for $\eps_{\rm min}/W= 0.1$, $E_0/W = -0.2$, and $U/W=1.0$. } 
  \label{fig:PhaseDiag01}
\end{figure}
%%%%%%%%%%%%%%%%%%%%%%%%%%%%%%%%%%%%

%%%%%%%%%%%%%%%%%%%%%%%%%%%%%%%%%%%%% 
\begin{figure}[t] 
  \begin{center}
    \includegraphics[width=7.5cm]{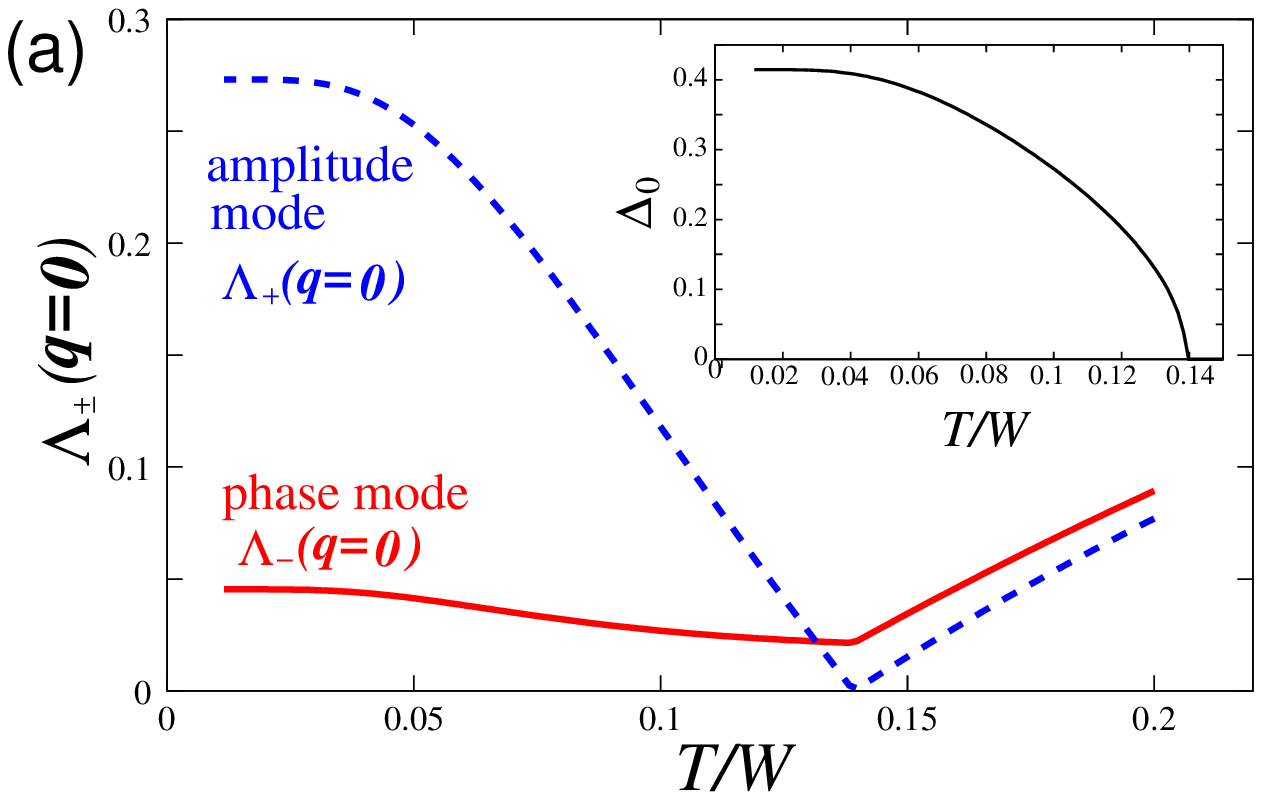}
    \includegraphics[width=7.5cm]{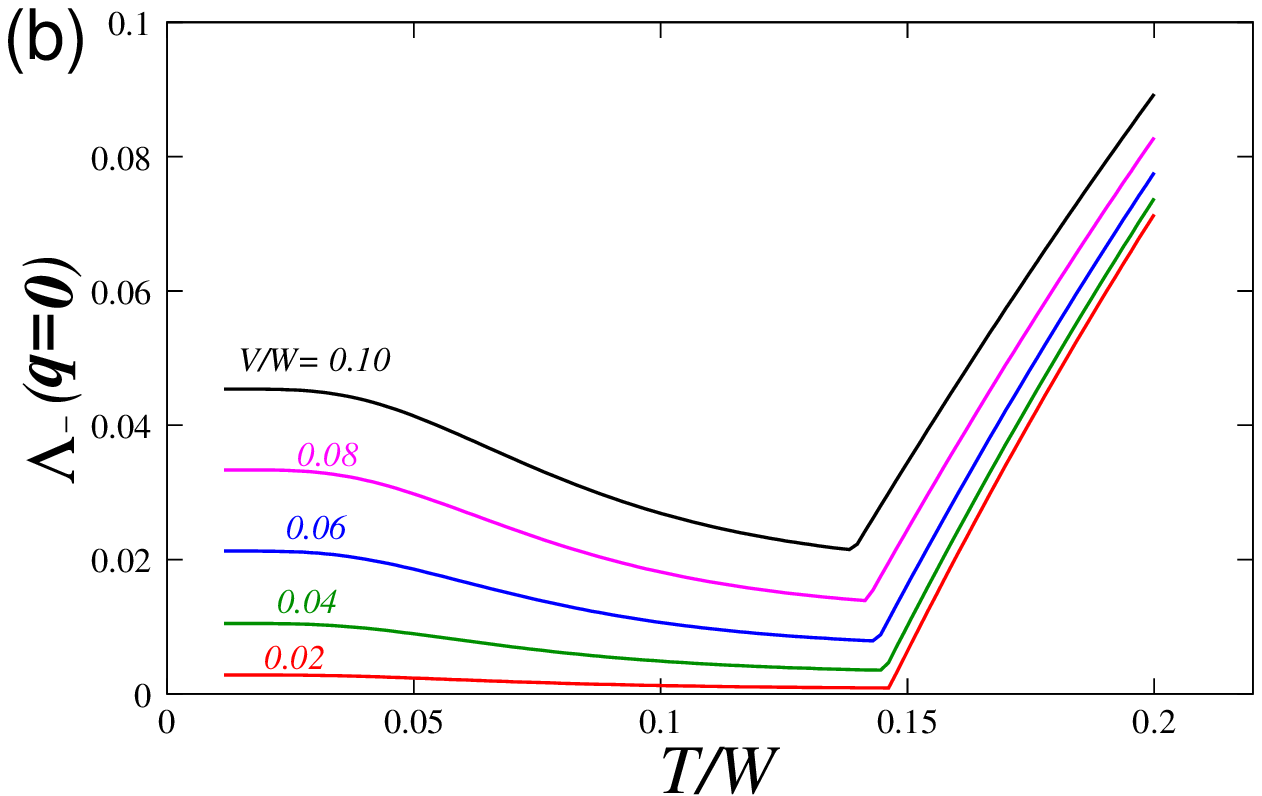}
    \includegraphics[width=7.5cm]{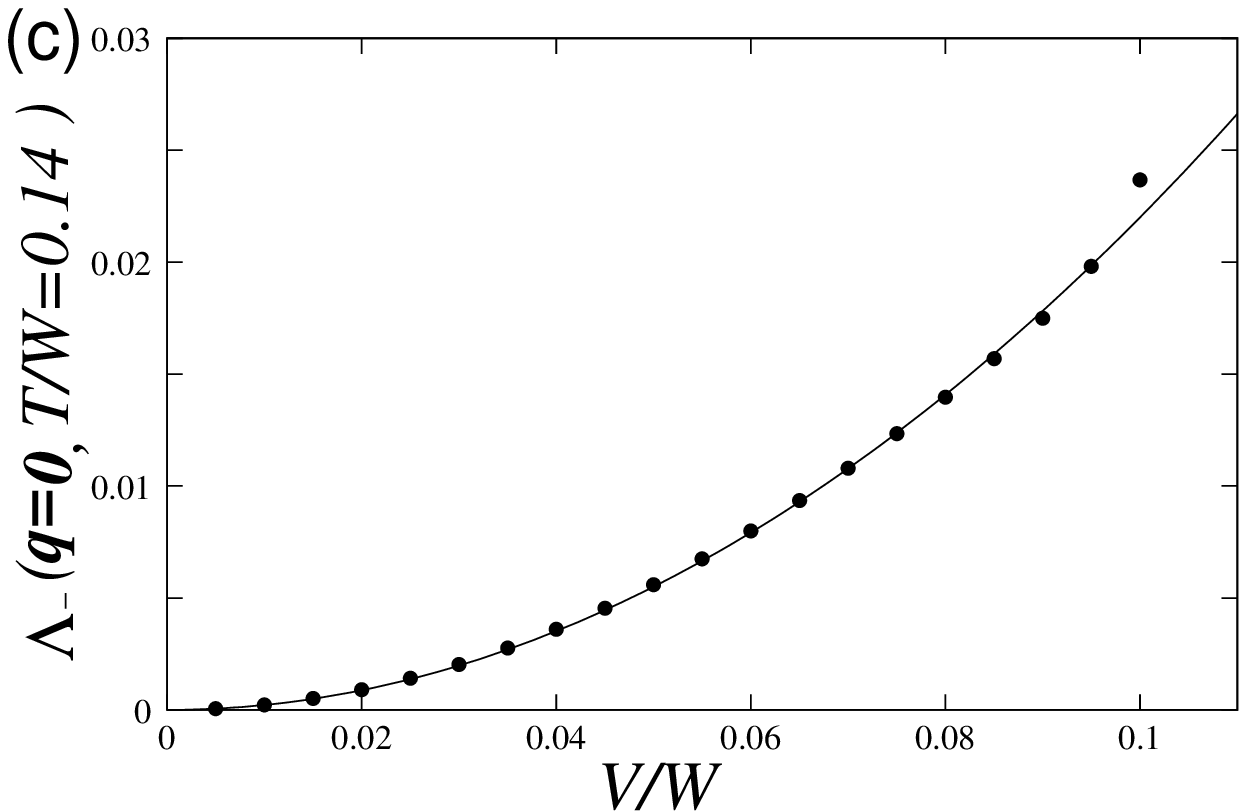} 
  \end{center} 
  \caption{(a) Temperature dependence of $\Lambda_\pm (\bmq= {\bm 0})$ for $\eps_{\rm min}/W= 0.1$, $E_0/W= -0.2$, $V/W=0.1$, and $U/W=1.0$. Inset: Temperature dependence of the order parameter $\Delta_0$ for the same parameters. (b) Temperature dependence of $\Lambda_- (\bmq= {\bm 0})$ for $\eps_{\rm min}/W= 0.1$, $E_0/W= -0.2$, $U/W=1.0$, and for several choices of $V/W$. (c) $\Lambda_-(\bmq= {\bm 0})$ as a function of $V$ at a fixed temperature of $T/W=0.14$. Dots are the calculated results, and the solid line represents a quadratic fit.} 
  \label{fig:CollectMode01}
\end{figure}
%%%%%%%%%%%%%%%%%%%%%%%%%%%%%%%%%%%% 

Figure~\ref{fig:PhaseDiag01}(a) shows the $T$-$E_0$ phase diagram for $\eps_{\rm min}/W= 0.1$, $V/W= 0.1$, and $U/W=1.0$. This phase diagram is drawn from the calculation of the GL coefficients $a_0$ and $b_0$, by changing the $f$-level position $E_0$ while keeping the $d$-electron band position unchanged. In the phase diagram, we see that a larger part of the phase transition is of the second order, but when $E_0$ touches and crosses the chemical potential, the transition becomes a first-order one. The inset of Fig.~\ref{fig:PhaseDiag01}(a) shows the temperature dependence of the bare correlation length,
%%%
\begin{equation}
  \xi^2 = \xi_1^2+ \xi_2^2+ \xi_3^2,
  \label{eq:xixi} 
\end{equation}
%%%
for $\eps_{\rm min}/W= 0.1$, $V/W= 0.1$, $E_0/W= -0.2$, and $U/W=1.0$, where $\xi_1^2$, $\xi_2^2$, and $\xi_3^2$ are defined in Eqs.~(\ref{eq:xixi01})-(\ref{eq:xixi03}). The sign of $\xi^2$ is always positive, which means that the gradient energy is positive and the mean-field solution is indeed a spatially uniform state, $\Delta_0$. The smallness of $\xi^2$ suggests that the fluctuation effects are important in the vicinity of the phase transition. In Fig.~\ref{fig:PhaseDiag01}(b), we show the $T$-$V$ phase diagram for $\eps_{\rm min}/W= 0.1$, $E_0/W= -0.2$, and $U/W=1.0$. We see that the ordered state is limited within a region of small $V$ values.

In Fig.~\ref{fig:CollectMode01}(a), we show the temperature dependence of the eigenvalue $\Lambda_\pm$ [Eq.~(\ref{eq:Lambda01})] in the $\bmq={\bm 0}$ limit, i.e., $\Lambda_\pm (\bmq= {\bm 0})$, for $\eps_{\rm min}/W= 0.1$, $E_0/W= -0.2$, $V/W=0.1$, and $U/W=1.0$. The minus branch is the phase mode, whereas the plus branch is the amplitude mode~\cite{Nagaosa-text}. As discussed in Refs.~\cite{Murakami20,Kaneko21}, the phase mode $\Lambda_-(\bmq= {\bm 0})$ has a non-zero value for a finite value of the hybridization $V$, which is however expected to vanish in the limit of $V \to 0$. In the inset of Fig.~\ref{fig:CollectMode01}(a), we plot the temperature dependence of the order parameter $\Delta_0$ [Eq.~(\ref{eq:Delta0})]. In Fig.~\ref{fig:CollectMode01}(b), we calculate the phase mode gap $\Lambda_-(\bmq= {\bm 0})$ by varying $V$ values. We see that, as expected, $\Lambda_-(\bmq= {\bm 0})$ is shrinking upon the decrease of $V$. This behavior is best seen in Fig.~\ref{fig:CollectMode01}(c), where the $\Lambda_-(\bmq= {\bm 0})$ value at $T/W=0.14$ is plotted as a function of $V/W$. The result shows that $\Lambda_-(\bmq= {\bm 0})$ is proportional to $V^2$, and the phase mode becomes the Goldstone mode in the $V \to 0$ limit.

%%%%%%%%%%%%%%%%%%%%%%%%%%%%%%%%%%%%%%%%%%%%%%%%
\section{Electric polarization and polarization current \label{Sec:V}} 
%%%%%%%%%%%%%%%%%%%%%%%%%%%%%%%%%%%%%%%%%%%%%%%

In this section, we first identify the polarization in our model by examining the response of the system to an external electric field. Then, using the knowledge~\cite{Dominicis75} that the GL action derived in the previous section has a one-to-one correspondence to the time-dependent GL theory, we investigate the low-energy dynamics of polarization and polarization current.

%%%%%%%%%%%%%%%%%%%%%%%%%%%%%%%%%%%%
\subsection{Electric polarization}
%%%%%%%%%%%%%%%%%%%%%%%%%%%%%%%%%%%
First, let us identify the polarization vector $\bmP$ by examining the effects of a static electric field $\bmE$ on the GL action. We consider the Hamiltonian describing the coupling between electrons and the external electric field, 
%%%
\begin{equation}
  {\cal H}_{\rm ext} = -|e| \sum_{\bmK}  \sum_\bmp \phi_\bmK
  \big( d^\dag_{\bmp+\bmK} d_\bmp + f^\dag_{\bmp+\bmK} f_\bmp \big) , 
\end{equation}
%%%
where $\phi_\bmK$ is the Fourier transform of the scalar potential giving $\bmE= - {\bm \nabla} \phi$. As in Ref.~\cite{Takayama70}, since we are interested in terms linear in $\phi$, we assume a single Fourier component of the scalar field and use the equation 
%%%
\begin{equation}
  \bmE= -\ui \bmK \phi_\bmK.
  \label{eq:E01}
\end{equation}
%%%

In the absence of the electric field, there is no term linear in $\Delta$ in the expansion of Eq.~(\ref{eq:S04}), i.e., ${\cal S}_1=0$. By contrast, in the presence of the electric field, there appears a term linear in both $\Delta$ and $\phi$, which reads 
%%%
\begin{equation}
  {\cal S}_1 = {\cal S}_{1a} + {\cal S}_{1b}+ {\cal S}_{1c}+ {\cal S}_{1d}, 
\end{equation}
%%%
where
%%%
\begin{eqnarray}
  {\cal S}_{1a} &=& -|e| \sqrt{\frac{T}{N}}  \sum_{\mathsf{q}, \mathsf{p}} \Delta^*_{\mathsf{q}} 
  G_1(\mathsf{p}+ \mathsf{K}) \phi_\bmK G_2(\mathsf{p}), \\
  {\cal S}_{1b} &=& -|e| \sqrt{\frac{T}{N}}  \sum_{\mathsf{q}, \mathsf{p}} \Delta^*_{\mathsf{q}} 
  G_2(\mathsf{p}+ \mathsf{K}) \phi_\bmK G_4(\mathsf{p}), \\
  {\cal S}_{1c} &=& -|e| \sqrt{\frac{T}{N}}  \sum_{\mathsf{q}, \mathsf{p}} \Delta_{\mathsf{q}} 
   G_1(\mathsf{p}- \mathsf{K}) \phi_\bmK G_3(\mathsf{p}), \\
   {\cal S}_{1d} &=& -|e| \sqrt{\frac{T}{N}}  \sum_{\mathsf{q},\mathsf{p}} \Delta_{\mathsf{q}} 
   G_3(\mathsf{p}-\mathsf{K}) \phi_\bmK G_4(\mathsf{p}),
\end{eqnarray}
%%%
and we introduced $\mathsf{K}=(\bmK,0)$ and disregard the dependence of the Green's function on the small wave number $\bmq$. Then, we use the gradient expansion used in Eqs.~(\ref{eq:xixi01})-(\ref{eq:xixi03}) above, and introduce the transformations $G_2 (\mathsf{p}) = \ui \widetilde{V}_\bmp G_0 (\mathsf{p})$ and $G_3 (\mathsf{p}) = -\ui \widetilde{V}_\bmp G_0 (\mathsf{p})$, where $\widetilde{V}_\bmp = -\ui V_\bmp  $ is a pure real number and 
%%%
\begin{equation}
  G_0 (\bmp, \ui \veps_n)= \frac{1}{D (\bmp, \ui \veps_n)} 
\end{equation}
%%%
with $D(\bmp, \ui \veps_n)$ being defined below Eq.~(\ref{eq:Gfunc01}). If we neglect the temporal fluctuations of $\Delta$, the action can be represented as 
%%%
\begin{equation}
  {\cal S}_1 = \frac{F_1}{T} 
\end{equation}
%%%
where the free energy $F_1$ linear in both $\Delta$ and $\phi$ can be written as 
%%%
\begin{equation}
  F_1 = -\sum_{\bmr_j} \bmP (\bmr_j) \cdot \bmE, 
  \label{eq:F1}
\end{equation}
%%%
and the polarization vector $\bmP$ is obtained, by using Eq.~(\ref{eq:E01}), as 
%%%
\begin{eqnarray}
  \bmP (\bmr_j) &=& {\bm \mu} \Big( \Delta (\bmr_j)+  \Delta^* (\bmr_j) \Big). 
  \label{eq:P01}
\end{eqnarray}
%%%
Here, $\Delta(\bmr_j)= \frac{1}{\sqrt{N}} \sum_\bmq \Delta_\bmq e^{\ui \bmq \cdot \bmr_j}$, and the dipole matrix element ${\bm \mu}$ is given by 
%%%
\begin{eqnarray}
  {\bm \mu} &=& {\bm \mu}_a + {\bm \mu}_b,
  \label{eq:mu01}
\end{eqnarray}
%%%
and
%%%
\begin{eqnarray} 
  {\bm \mu}_a &=& -|e| \frac{T}{N}  \sum_{\mathsf{p}}
  \big( G_1(\mathsf{p}) \big)^2 G_0(\mathsf{p}) \widetilde{V}_p
  (\bmv_\bmp \cdot {\bf \hat{e}}) {\bf \hat{e}},
  \label{eq:mu01a}\\ 
  {\bm \mu}_b &=& -|e| \frac{T}{N}  \sum_{\mathsf{p}}
  G_0(\mathsf{p}) G_1 (\mathsf{p})G_4 (\mathsf{p}) \widetilde{V}_p
  (\bmv_\bmp \cdot {\bf \hat{e}}) {\bf \hat{e}},
  \label{eq:mu01b}
\end{eqnarray}
%%%
where ${\bf \hat{e}}= \bmE/E$ is the unit vector along $\bmE$.

From these equations, we see that the polarization vector $\bmP$ and the dipole matrix element ${\bm \mu}$ shrink in the limit of the vanishing hybridization, $V \to 0$. Conversely, in the presence of the hybridization, we obtain a non-zero $\bmP$ in the excitonic insulator phase. These quantities, $\bmP$ and ${\bm \mu}$, are unambiguously determined from our model Hamiltonian in the present work, whereas ${\bm \mu}$ was introduced by hand as a phenomenological parameter in Ref.~\cite{Portengen96b}.

%%%%%%%%%%%%%%%%%%%%%%%%%%%%%%%%%%%%%%
\subsection{Polarization current}
%%%%%%%%%%%%%%%%%%%%%%%%%%%%%%%%%%%%%%

Since the relationship between the polarization $\bmP$ and the order-parameter field $\Delta$ is established in Eq.~(\ref{eq:P01}), let us now discuss the dynamics of the fluctuations of $\Delta$ as well as that of the polarization $\bmP$. 

In the following, we assume that a weak electric field is applied along the $z$-axis to align the polarization, i.e., $\bmE= E {\bf \hat{z}}$. Then, we decompose the polarization into the mean-field value and fluctuations, 
%%%
\begin{equation}
  P^z (\bmr) = P_0 + \delta P^z (\bmr),
  \label{eq:Pz01}
\end{equation}
%%%
where we used the symbol $\bmr$ to denote the real space position instead of its lattice form $\bmr_j$. Using Eq.~(\ref{eq:P01}), we can relate the right-hand side of Eq.~(\ref{eq:Pz01}) with the order parameter $\Delta$. The first term on the right-hand side, $P_0= 2 \mu^z \Delta_0$, is the spatially uniform polarization with $\mu^z$ being the $z$-component of the inter-band dipole matrix element ${\bm \mu}$. The second term, $\delta P^z (\bmr)$, is represented in the wave number space as $\delta P^z_\bmq = \mu^z ( \delta \Delta_\bmq+ \delta \Delta^*_{-\bmq} )$, which, by using Eq.~(\ref{eq:unitaryT02}), is transformed to 
%%%
\begin{equation}
  \delta P^z_\bmq = \mu^z \sqrt{2} \beta_\bmq, 
  \label{eq:P02}
\end{equation}
%%%
where $\beta_\bmq$ is the amplitude mode defined by Eq.~(\ref{eq:unitaryT02}). 

From Eq.~(\ref{eq:P02}), we see that in order to understand the dynamics of $\delta P^z$, we need to examine the dynamics of $\beta_\bmq$, or the amplitude mode of the Gaussian fluctuations of $\Delta$. For this purpose, we first define the effective Hamiltonian ${\cal H}_{\rm GL}$~\cite{Landau-text} that describes the fluctuations of $\Delta$: 
%%%
\begin{equation}
  {\cal S}_{\rm GL}|_{\text{static limit}} = \frac{{\cal H}_{\rm GL}}{T}, 
\end{equation}
where the ``static limit'' means to neglect the temporal fluctuations of $\Delta$~\cite{Altland-Simons}. The resultant effective Hamiltonian ${\cal H}_{\rm GL}$ can be diagonalized in the same manner as ${\cal S}_{\rm GL}$ by introducing two fields $\alpha_\bmq$ and $\beta_\bmq$, the result of which is given by 
%%%
\begin{equation}
  {\cal H}_{\rm GL} = {\cal F}_{\rm MFA} 
  + \sum_{\bmq} \Big( \Lambda_- (\bmq) \alpha^*_{\bmq} \alpha_{\bmq} 
  + \Lambda_+ (\bmq) \beta^*_{\bmq} \beta_{\bmq} \Big), 
\end{equation}
%%%
where $\Lambda_\pm (\bmq)$ is defined by Eq.~(\ref{eq:Lambda01}). 

Next, we use the idea~\cite{Dominicis75,Larkin-text} that the dynamic equation derived from the stationary phase approximation of the quantum GL action [Eq.~(\ref{eq:S_GL02})] is equivalent to the following time-dependent GL equations~\cite{Hohenberg77}: 
%%%
\begin{eqnarray}
  \frac{\partial}{\partial t} \alpha_\bmq 
  &=& - \Gamma_- (\bmq) \frac{\partial \; {\cal H}_{\rm GL}}{\partial \, \alpha^*_\bmq }
  + \zeta_-(\bmq,t), \label{eq:alpha01} \\ 
%  =  - \gamma_-(\bmq) \Lambda_- (\bmq) \alpha_\bmq , 
  \frac{\partial}{\partial t} \beta_\bmq  
  &=& - \Gamma_+ (\bmq) \frac{\partial \; {\cal H}_{\rm GL}}{\partial \, \beta^*_\bmq }
  + \zeta_+(\bmq,t), \label{eq:beta01}
  %  = - \gamma_+(\bmq) \Lambda_+(\bmq) \beta_\bmq. 
\end{eqnarray}
%%%
where $\Gamma_\pm (\bmq)= 1/(\Gamma_1^{-1}(\bmq) \pm 2 \Gamma_2^{-1}(\bmq))$ [see Eq.~(\ref{eq:Omega01})], and $\zeta_\pm(\bmq,t)$ is the thermal noise field with zero mean and variance proportional to the temperature.

Now we discuss the form of the damping coefficient $\Gamma_\pm$. If we used the microscopic expression of the damping coefficient $\Gamma_i$ ($i= 1$, $2$, and $3$) [Eqs.~(\ref{eq:Gamma01})-(\ref{eq:Gamma03})] for a pure system without any imperfections, we would obtain the Landau damping form, $\Gamma_i \sim q$~\cite{Hertz76,Adachi09,Klein18}. However, as for the transport phenomena in the long-wavelength low-frequency limit, we expect that the transport is dominated by the diffusive one in a real material. Therefore, we assume the following diffusive form for $\Gamma_\pm$~\cite{Hertz76,Adachi18} 
%%%
\begin{equation}
  \Gamma_\pm(\bmq) = \Gamma_\pm(0) + D_\pm q^2, 
\end{equation}
%%%
where $D_\pm$ plays the role of the bare diffusion coefficients for $\beta$ and $\alpha$ fields, respectively, and we assume $\Gamma_\pm(0)$ to be non-zero because there is no physical reason for the conservation of $P^z$~\cite{Klein18}. 

Then, using Eqs.~(\ref{eq:beta01}) and (\ref{eq:P02}), we obtain the dynamic equation for the polarization $\delta P^z_\bmq$: 
%%%
\begin{equation}
  \left( \frac{\partial}{\partial t}
  + {\cal D}_+^{\rm eff} q^2 + \frac{1}{\tau^{\rm eff}_+} \right) \delta P_\bmq^z = 0, 
\end{equation}
%%%
where ${\cal D}_+^{\rm eff}= D_+ \Lambda_+(0)$ and $1/{\tau_+^{\rm eff}} = \Gamma_+(0) \Lambda_+(0)$. Going back to the real space representation, the above equation can be transformed into the continuity equation~\cite{Bennett65} 
%%%
\begin{equation}
  \frac{\partial}{\partial t} \delta P^z (\bmr) 
  + {\rm div} {\bm J}_P + \frac{1}{\tau^{\rm eff}_+}  \delta P^z (\bmr) = 0,
  \label{eq:Pdiffeq01}
\end{equation}
%%%
where the polarization current is given by
%%%
\begin{equation}
  {\bm J}_P = - {\cal D}_{+}^{\rm eff} {\bm \nabla} \delta P^z (\bmr),
  \label{eq:Jp01}
\end{equation}
%%%
and the polarization fluctuation is described by the $\beta$ field [Eq.~(\ref{eq:P02})].

The final result, Eq.~(\ref{eq:Jp01}), shows that we can define the polarization current purely electronically without recourse to the lattice degrees of freedom. Note that the polarization diffusion equation derived above for electronic ferroelectricity has the same form as that for displacive ferroelectricity~\cite{Bauer22}. Note also that in both cases the polarization current, or ferron current, carries heat. In the case of displacive ferroelectricity, heat conduction carried by ferron excitations has recently been demonstrated~\cite{Wooten23}. 

%%%%%%%%%%%%%%%%%%%%%%%%%%%%%%%%%%%%%%%%
\section{Discussion and Conclusion \label{Sec:VI}} 
%%%%%%%%%%%%%%%%%%%%%%%%%%%%%%%%%%%%%%%%
In this paper, starting from the excitonic insulator model of electronic ferroelectricity~\cite{Batyev80,Portengen96a,Portengen96b}, Eq.~(\ref{eq:H_exciton01}), we have examined the low-energy dynamics and transport of the electric polarization. To this end, we have employed the functional integral technique to calculate the GL action [Eq.~(\ref{eq:S_GL02})] as well as constructed the corresponding time-dependent GL equations for the order parameter fluctuations [Eqs.~(\ref{eq:alpha01}) and (\ref{eq:beta01})]. 

The excitonic insulator phase considered in this work is characterized by a complex scalar order parameter. Indeed, noticing that the $\zeta$ field in Eq.~(\ref{eq:Stratnovich01}) has a Gaussian distribution with zero mean, the expectation value of the order parameter is given by 
%%%
\begin{equation}
  \langle \Delta_\bmq \rangle = - \frac{U}{\sqrt{N}}
  \sum_\bmp \langle f^\dag_\bmp d_{\bmp+ \bmq} \rangle, 
  \label{eq:Stratnovich02}
\end{equation}
%%%
where the phase of the order parameter $\Delta_\bmq$ corresponds to the phase difference between the $d$-electron wave function and the $f$-electron wave function. It is important to note here that, as is clear from Eqs.~(\ref{eq:mu01a}) and (\ref{eq:mu01b}), for the excitonic insulator phase to simultaneously possess the electronic ferroelectricity, we need to assume the presence of a non-zero hybridization $V_\bmp$.

The Ginzburg-Landau approach used in this work is justified near the transition temperature $T_{\rm c}$, and if the band gap between $d$- and $f$-electrons is comparable to the thermal energy of $T_{\rm c}$, the polarization of electronic ferroelectricity may be screened by surface charge accumulation of thermally excited carries. In Fig.~\ref{fig:PhaseDiag01}(a), this could be the case for the $f$-level energy being located around $E_0/W \approx -0.2$, giving the highest $T_{\rm c}$ in the ferroelectric dome (recall that the bottom of the $d$-band, $\eps_{\rm min}$, is always fixed at $\eps_{\rm min}/W= 0.1$ in our calculation as in Fig.~\ref{fig:DOS01}). Conversely, when the $f$-level energy is located around $E_0/W \approx -0.4$ giving much lower $T_{\rm c}$, the thermal energy of $T_{\rm c}$ is much smaller than the band gap, such that the effects of thermally excited carries are negligible. Therefore, strictly speaking, the concept of electronic ferroelectricity in the present model is well-defined at the lower $T_{\rm c}$ side of the ferroelectric dome in Fig.~\ref{fig:PhaseDiag01}(a).

One of our motivations in this work is to formulate the polarization transport in the electronic ferroelectricity without relying on the phonon degrees of freedom, which is in contrast to the case of displacive ferroelectricity where the polarization transport is formulated by using phonon operators~\cite{Tang22a,Tang22b}. In this regard, we have clarified the following points. First, reflecting the above-mentioned fact that the ferroelectric phase is characterized by a scalar order parameter in the present model, we have found that the electric polarization has only the longitudinal dynamics. Second, the longitudinal fluctuation of the polarization density, $\delta P^z(\bmr)$, gives rise to the polarization diffusion, which is described by Eq.~(\ref{eq:Pdiffeq01}). Then, the polarization current [Eq.~(\ref{eq:Jp01})] is defined in terms of the electronic ferroelectricity order parameter given by Eq.~(\ref{eq:P01}). Therefore, in the present system of electronic ferroelectricity, we have succeeded in describing the transport of polarization in such a way that it is free from the issue of subtle distinction between phonon excitations and ``ferron'' excitations~\cite{Tang22a,Tang22b}. Because of its electronic nature, the time scale of ferron excitations in the electronic ferroelectricity is expected to be much faster than that in the displacive ferroelectricity. However, precise analysis of such a difference is left for future studies.

To conclude, starting from the excitonic insulator model of electronic ferroelectricity, we have theoretically investigated the low-energy dynamics and transport of electric polarization. Based on the GL action derived microscopically, we have revealed that the longitudinal fluctuations are relevant to the transport of electric polarization, and we have constructed the polarization diffusion equation. Although experimental results are unavailable at present, we hope the present result gives a constructive input to future experiments. 

%%%
\acknowledgments 
%%%
We are grateful to K. Uchida, R. Iguchi, J. Otsuki, and N. Yokoi for useful comments and discussions. H.A. was financially supported by JSPS KAKENHI Grant No. 22H01941 and the Asahi Glass Foundation, N.I. was supported by JSPS KAKENHI Grant No. 22H01942; and E.S. was supported by JSPS KAKENHI Grant No. JP19H05600, JST CREST Grants No. JPMJCR20C1 and No. JPMJCR20T2, Integrated Circuits Centers (X-NICS, Grant No. JPJ011438), Institute for AI and Beyond of the University of Tokyo, and IBM-UTokyo Laboratory. 

%\appendix


%\clearpage

%\subsection*{Figure Captions} 

\end{document}